# Unified Friction Formulation from Laminar to Fully Rough Turbulent Flow

**Dejan Brkić** [1,*] **and Pavel Praks** [1,2,*]

[1] European Commission, Joint Research Centre (JRC), Directorate C - Energy, Transport and Climate, Unit C3: Energy Security, Distribution and Markets, Via Enrico Fermi 2749, 21027 Ispra (VA), Italy
[2] IT4Innovations National Supercomputing Center, VŠB—Technical University of Ostrava, 17. listopadu 2172/15, 708 00 Ostrava, Czech Republic
* Correspondence: dejanbrkic0611@gmail.com (D.B.); Pavel.Praks@gmail.com (P.P.)

**Abstract:** This paper gives a new unified formula for the Newtonian fluids valid for all pipe flow regimes from laminar to the fully rough turbulent. It includes laminar, unstable sharp jump from laminar to turbulent, and all types of the turbulent regimes: smooth turbulent regime, partial non-fully developed turbulent and fully developed rough turbulent regime. The formula follows the inflectional form of curves as suggested in Nikuradse's experiment rather than monotonic shape proposed by Colebrook and White. The composition of the proposed unified formula consists of switching functions and of the interchangeable formulas for laminar, smooth turbulent and fully rough turbulent flow. The proposed switching functions provide a smooth and a computationally cheap transition among hydraulic regimes. Thus, the here presented formulation represents a coherent hydraulic model suitable for engineering use. The model is compared to existing literature models, and shows smooth and computationally cheap transitions among hydraulic regimes.

**Keywords:** Turbulent flow; Laminar flow; Pipes; Friction factor; Hydraulics; Monotonic roughness, Inflectional roughness; Smooth curve contact; Moody diagram; Hydraulic resistance.

## 1. Introduction

In hydraulics, resistance in pipe flow is represented usually through the Darcy flow friction factor $\lambda$ which depends on the Reynolds number Re and the relative roughness of inner pipe surface $\varepsilon$ [1]. All three quantities are dimensionless. For pipe flow, the Reynolds number Re usually takes values between 0 and $10^8$, while the relative roughness of inner pipe surface $\varepsilon$ from 0 to 0.05. The relative roughness of inner pipe surface is not only characteristic of pipe material and its condition, but depends also on hydraulic flow regime ruled by the thickness of thin laminar boundary sub-layer of fluid near inner pipe surface [2,3] (Figure 1-up). In general, during pipe flow few different hydraulic regimes can occur (Figure 1); laminar $(a)$, sharp transition from laminar to the smooth turbulent $(b)$, smooth turbulent $(c_1)$, non-fully developed partially turbulent $(c_2)$ and fully developed rough turbulent $(c_3)$:

$(a)$: Absence of vortices is characteristic of laminar regime, while roughness of inner pipe surface does not have effect on flow,

$(b)$ Transition for laminar to turbulent regime is sharp and almost immediate and can be described with sudden increase of the friction factor $\lambda$,

$(c)$ Higher values of the Reynolds number Re with vortices in flow are characteristic of turbulent regime: $(c_1)$ First vortices in the middle of pipe are characteristic for the smooth turbulent regime, while still roughness of inner pipe surface is covered with laminar sub-layer, $(c_2)$ With increase of the Reynolds number Re, thickness of the laminar sub-layer decreases and roughness of inner pipe surface stars to have important role, $(c_3)$. In the case of fully rough turbulent flow, the roughness of inner pipe surface takes the dominant role.

Physical description of the different hydraulic regimes is given in Figure 1-up, while related diagrams in Figure 1-down.

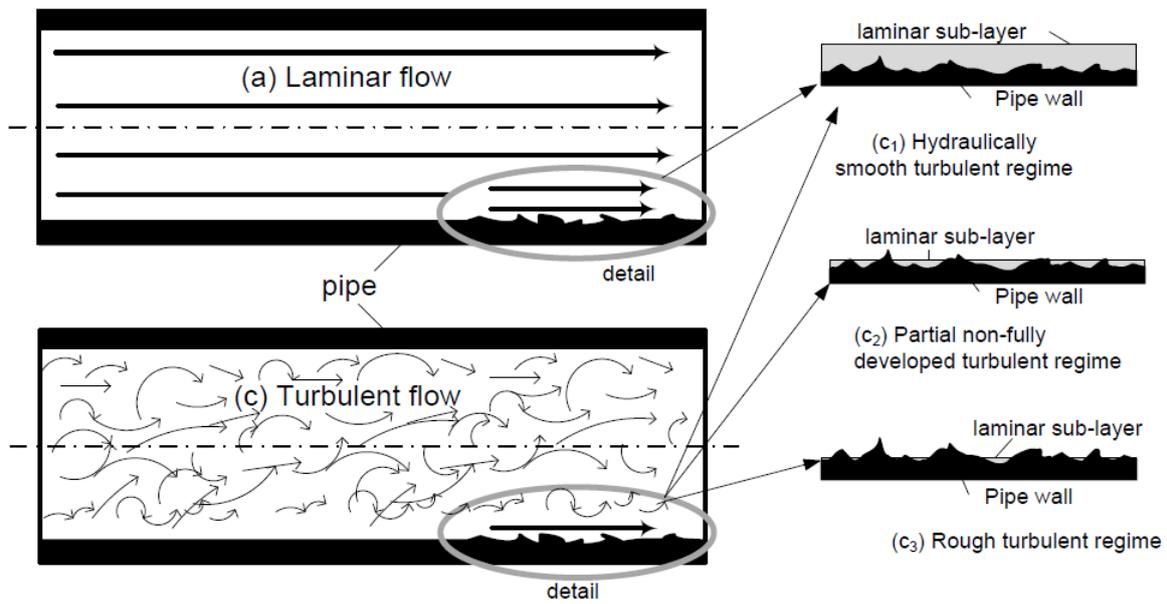

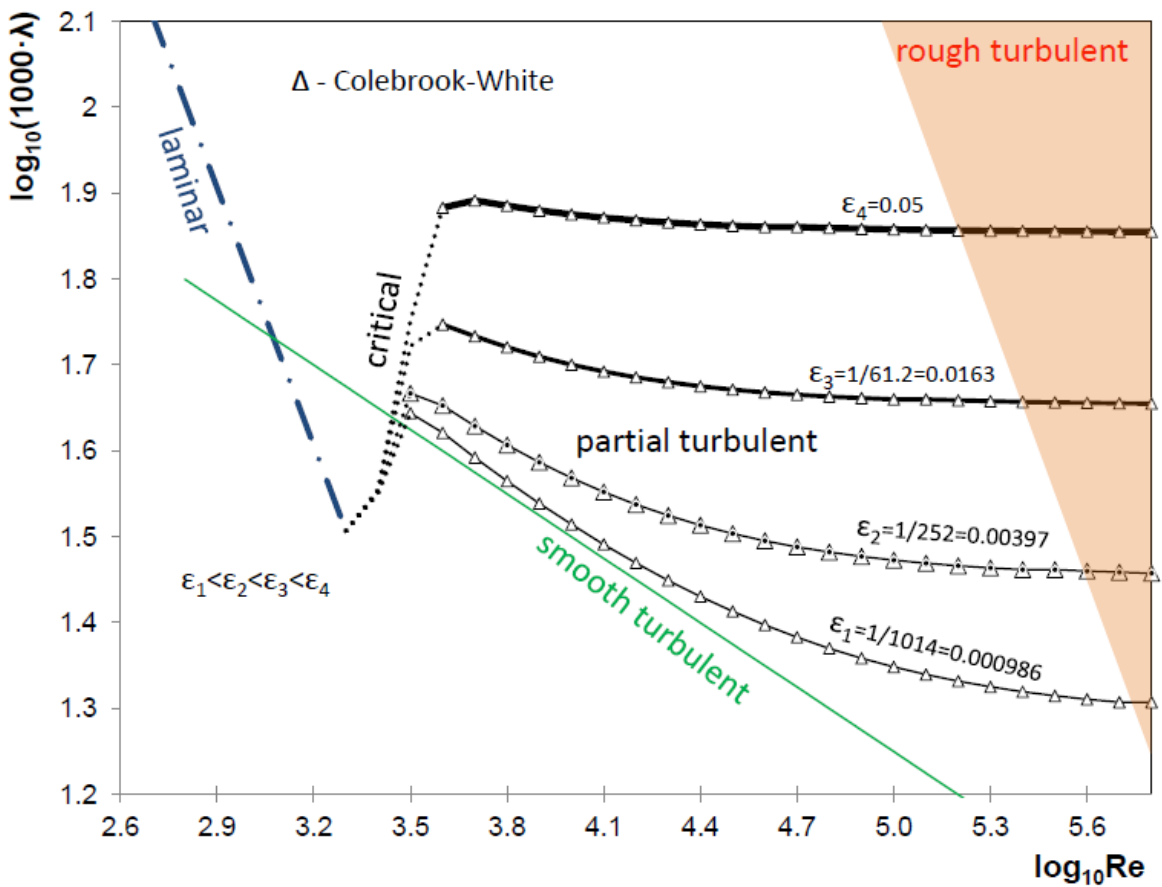

**Figure 1.** Hydraulic regimes for pipe flow: laminar $(a)$, sharp transition from laminar to the smooth turbulent (critical) $(b)$, smooth turbulent $(c_1)$, non-fully developed partially turbulent $(c_2)$ and fully developed rough turbulent $(c_3)$; physical interpretation –up, and diagrams - down

Similar as Figure 1, the good representation of different hydraulic regimes gives widely known Moody diagram [4-6]. Turbulent part of the Moody diagram is based on the Colebrook equation [7].

While the formula for laminar regime; is very well theoretically founded, the formulas for all types of turbulent regimes are empirical [8-10]. Although coherent unified models for all hydraulic regimes exist [11-13], they are not flexible as here presented because they are fixed (they do not allow changing separate formulas valid for different hydraulic regimes). In contrary, the here presented unified formula allows any of the formulas for the particular hydraulic regime to be included: one for laminar ($a$), one for hydraulically smooth turbulent ($c_1$) and one for fully developed rough turbulent ($c_3$). Thus, the here presented unified formula provides a smooth transition among hydraulic regimes using three switching functions $y_1$, $y_2$ and $y_3$; see Eqs. (7-9) of this paper. The proposed unified formula is given by the following composition, Eq. (1):

$$\lambda = (1 - y_1) \cdot (a) + (y_1 - y_3) \cdot (c_1) + y_2 \cdot (c_3) \tag{1}$$

In Eq. (1), $y_1$, $y_2$ and $y_3$ represent switching functions; ($a$) is a formula for laminar flow, $\lambda = \frac{64}{Re}$; while ($c_1$) is for smooth turbulent flow, $\lambda = \zeta(Re)$; and finally ($c_3$) is for fully developed rough turbulent flow, $\lambda = \varsigma(\varepsilon)$. Here Re is the Reynolds number, $\varepsilon$ is relative roughness of inner pipe surface while $\zeta$ and $\varsigma$ are functional symbols, respectively. Functions $\zeta$ and $\varsigma$ are empirical and related equations are available from literature [14].

The proposed unified formula follows the inflectional shape of curve as suggested in Nikuradse's experiment [8,15] rather than the monotonic shape of curves as proposed by Colebrook and White [7,9]. In our case, such shape is provided through carefully fitting of the switching functions $y_1$, $y_2$ and $y_3$. The Nikurdse's inflectional shape of curves is confirmed also by the recent Princeton and Oregon experiments with flow friction [14]. These two devices are significantly different but they reached similar conclusions; the Princeton facility uses compressed air, while Oregon uses helium, oxygen, nitrogen, carbon dioxide, and sulfur hexafluoride; the Princeton device weights approximately 25 tons, while the Oregon device weighs approximately 0.7 kg. Variant of the Moody diagram with included the Nikurdse's inflectional shape of curves also exists [16].

**2. Previous works and source of their differences**

Only few formulas valid for all hydraulic regimes exist; Eqs. (2-4), [11-13] respectively:

$$\lambda = 8 \cdot \sqrt[12]{\left(\frac{8}{Re}\right)^{12} + \left[\left(-2.457 \cdot \ln\left(\left(\frac{7}{Re}\right)^{0.9} + 0.27 \cdot \varepsilon\right)\right)^{16} + \left(\frac{37530}{Re}\right)^{16}\right]^{-1.5}} \tag{2}$$

$$\lambda = \sqrt[8]{\left(\frac{64}{Re}\right)^8 + 9.5 \cdot \left[\ln\left(\frac{\varepsilon}{3.7} + \frac{5.74}{Re^{0.9}}\right) - \left(\frac{2500}{Re}\right)^6\right]^{-16}} \tag{3}$$

$$\left.\begin{array}{l}\lambda = 0.11 \cdot \sqrt[4]{\frac{\lambda_1 + \varepsilon + (28 \cdot \lambda_1)^{14}}{1 + 115 \cdot (28 \cdot \lambda_1)^{10}}} \\ \lambda_1 = \frac{68}{Re}\end{array}\right\} \tag{4}$$

In Eqs. (2-4), $\lambda$ is the Darcy friction factor, Re is the Reynolds number Re, and $\varepsilon$ the relative roughness of inner pipe surface (all three quantities are dimensionless).

Turbulent part of Eqs. (2,3) are based on the Colebrook formula [7], while Eq. (4) is based on formulas from Russian engineering practice [17]. Turbulent part of Eq. (2) [11] is based on Churchill approximation of the Colebrook equation [18], while Eq. (3) [12] is based on Swamee and Jain approximation of the Colebrook equation [19]. As shown in Figure 2, the formulas from Russian practices and those based on the Colebrook equation give almost identical results for low values of relative roughness ε, while in the case of higher values of relative roughness ε, formulas from Russian practices give lower values for flow friction factor $\lambda$ compared with the Colebrook equation. Both the Colebrook equation and formulas from Russian practice have monotonic shape of flow friction curves (Figure 2), which is disputed by some new experiments such as those from Princeton or Oregon laboratories [15], but also by an experiment of Nikuradse [8]. The experiment of Nikuradse had been conducted in 1932 and 1933, while the experiment by Colebrook and White had been reported in 1937 [9]). The experiment of Nikuradse proposes a shape of curves in turbulent regime with an interval of declining values of friction factor λ before they reach their final maximal value for the fully developed rough turbulent flow. This property of flow friction is known as the Nikuradse's inflectional shape of transition to full turbulent regime (Figure 3) [20,21].

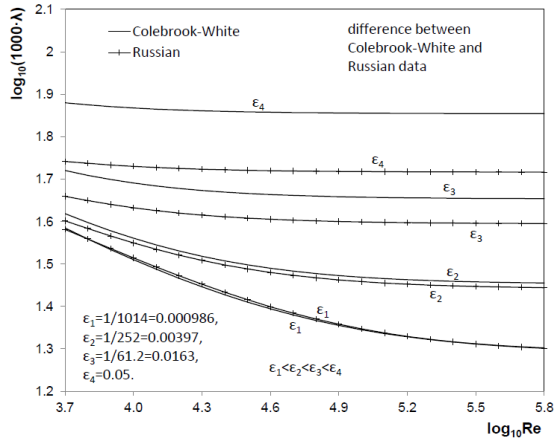 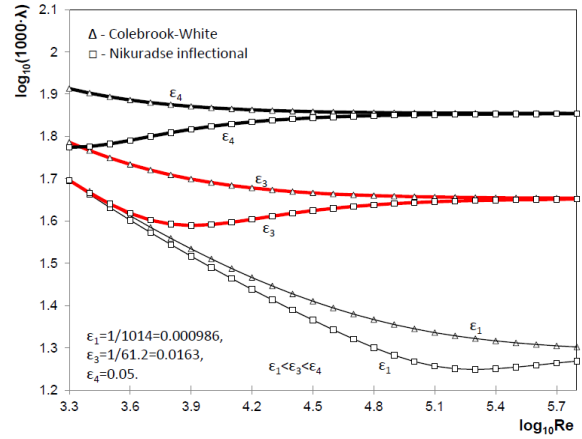

**Figure 2.** Different values of flow friction factor obtained using the Colebrook formula versus the formulas from Russian practice

**Figure 3.** Different shapes flow friction factor curves obtained using the Colebrook formula versus those that follow Nikuradse's inflectional law

None of the currently available formulas valid for all hydraulic regimes; Eqs. (2-4), do not follow Nikuradse's inflectional law. The Colebrook equation can be extended to fit the data from Nikuradse's experiment; Eq. (5) [16]:

$$\left. \begin{array}{l} (c_2) \sim \frac{1}{\sqrt{\lambda}} = -2 \cdot \log_{10}\left(\frac{2.51}{Re} \cdot \frac{1}{\sqrt{\lambda}} + \frac{\varepsilon}{3.71} \cdot e^E\right) \\ E = \frac{-31.13}{Re \cdot \varepsilon} \cdot \frac{1}{\sqrt{\lambda}} \end{array} \right\} \quad (5)$$

In Eq. (5), $e$ is exponential function, where $E=0$ gives the Colebrook equation without Nikuradse's extension. The Colebrook equation is given in implicit form in respect to the flow friction factor λ [22-25], while the formulas from Russian practice is mostly belong to the explicitly given power-law type [26]. Anyway, as shown in Figure 4, the Colebrook equation is developed to unify in a smooth asymptotic way the von Karman-Prandtl equations for smooth turbulent flow and for fully rough flow; Eq. (6), but always having in mind that $\log(\alpha) + \log(\beta) \neq \log(\alpha + \beta)$:

$$\left. \begin{array}{l} (c_1) \sim \frac{1}{\sqrt{\lambda}} = 2 \cdot \log_{10}(Re \cdot \sqrt{\lambda}) - 0.8 = -2 \cdot \log_{10}\left(\frac{2.51}{Re} \cdot \frac{1}{\sqrt{\lambda}}\right) \\ (c_3) \sim \frac{1}{\sqrt{\lambda}} = 1.14 - 2 \cdot \log_{10}(\varepsilon) = -2 \cdot \log_{10}\left(\frac{\varepsilon}{3.71}\right) \end{array} \right\} \rightarrow (c_2) \sim \frac{1}{\sqrt{\lambda}} = -2 \cdot \log_{10}\left(\frac{2.51}{Re} \cdot \frac{1}{\sqrt{\lambda}} + \frac{\varepsilon}{3.71}\right) \quad (6)$$

In Eq. (6), $(c_1)$ presents smooth turbulent flow, $(c_2)$ transitional non-fully developed turbulent flow and $(c_3)$ fully developed rough turbulent flow. Similar strategies to unify and to modify the equation to conform to the certain laws (in our case Nikuradse's inflectional law) are used for our unified equation; see Eq. (1). On the other hand, Colebrook [9]; see Eqs. (5) and (6), uses logarithmic law to unify $(c_1)$ and $(c_3)$ in $(c_2)$. In our approach, we use switching functions $y_1$, $y_2$ and $y_3$ to unify $(a)$, $(c_1)$ and $(c_3)$ in one coherent hydraulic model, where $(a)$ represents laminar regime as already previously explained by Eq. (1).

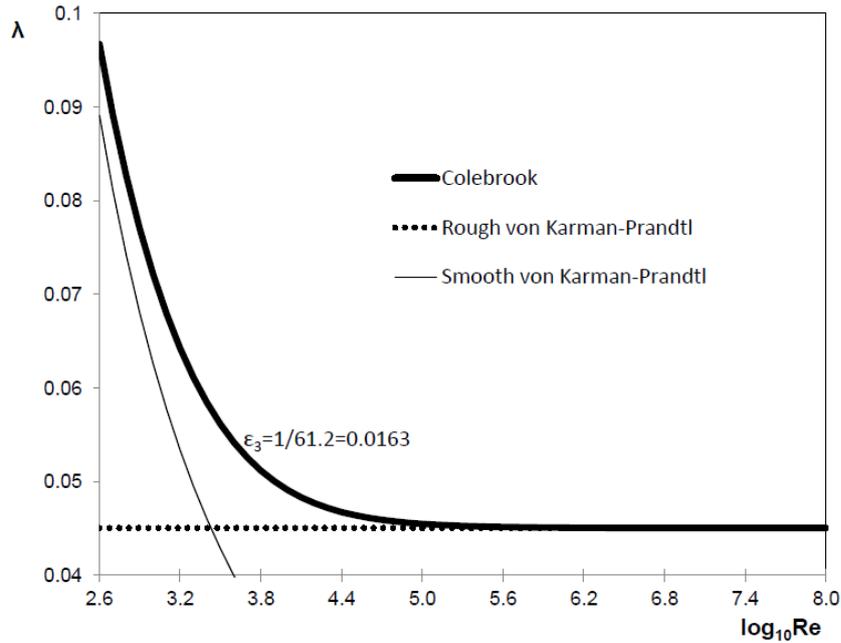

**Figure 4.** Von Karman-Prandtl equations for smooth turbulent flow and for fully rough flow unified by Colebrook using logarithmic function in one coherent hydraulic model

## 3. Switching functions, Friction factors, New formulation and Comparative analysis

Required inputs for the here presented unified formula; Eq. (1), are shown here: switching functions $y_1$, $y_2$ and $y_3$ and formulas for friction factors for the certain hydraulic zones: $(a)$-laminar, $(c_1)$-smooth turbulent flow, and $(c_3)$-fully developed rough turbulent flow.

*3.1. Switching functions*

The unified formula valid for all hydraulic cases, in our case depends directly on switching functions $y_1$, $y_2$ and $y_3$. They provide smooth transition between different hydraulic regimes. Connection of curves provided using separate formulas for $(a)$-laminar, $(c_1)$-smooth turbulent flow, and $(c_3)$-fully developed rough turbulent flow, is not smooth without help of these three carefully selected switching functions $y_1$, $y_2$ and $y_3$. These three functions are rational, and they are generated in HeuristicLab, a software environment for heuristic and evolutionary algorithms [27]. The proposed switching functions have a simple form and therefore can be incorporated easily in computer codes without any significant burden of the Central Processing Unit [28]. The proposed switching functions $y_1$, $y_2$ and $y_3$ are represented by Eqs. (7-9) and Figures 5-7, respectively.

$$y_1 = 1 - \frac{1048}{\frac{4.489}{10^{20}} \cdot Re^6 \cdot \left(0.148 \cdot Re - \frac{2.306 \cdot Re}{0.003133 \cdot Re + 9.646}\right) + 1050} \qquad (7)$$

$$y_2 = 1.012 - \frac{1}{0.02521 \cdot Re \cdot \varepsilon + 2.202} \qquad (8)$$

$$y_3 = 1 - \frac{1}{0.000389 \cdot Re^2 \cdot \varepsilon^2 + 0.0000239 \cdot Re + 1.61} \qquad (9)$$

In Eqs. (7-9), symbols $y_1$, $y_2$ and $y_3$ denote switching functions.

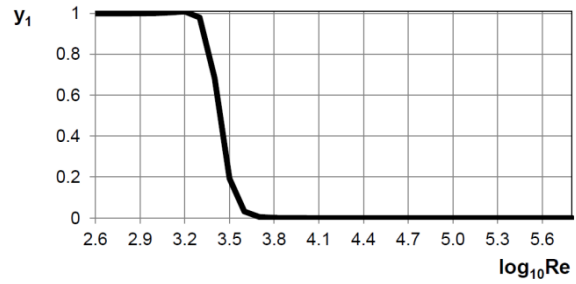

**Figure 5.** Switching function $y_1$ as a function of the Reynolds number with the main purpose to provide transition between $(a)$-laminar, $(c_1)$-smooth turbulent flow

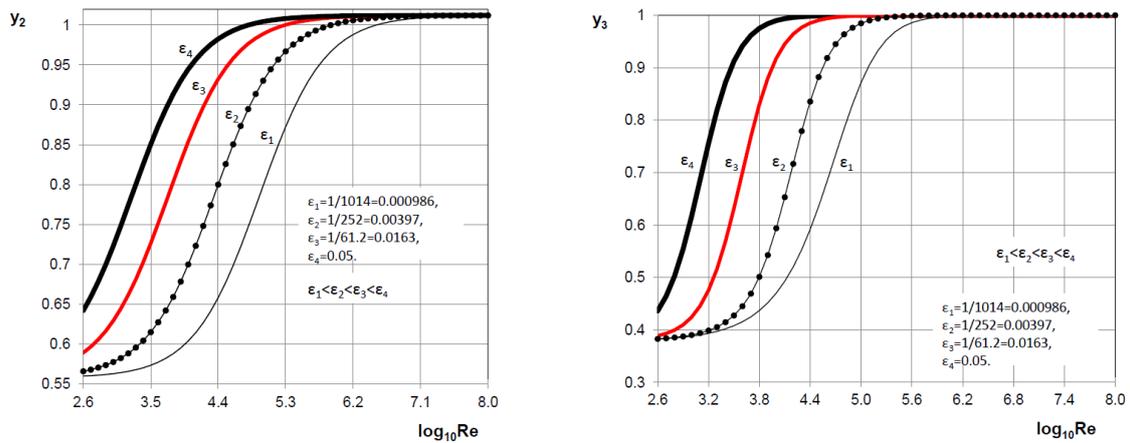

**Figure 6.** Switching function $y_2$ and $y_3$ with the main purpose to provide transition between $(c_1)$-smooth turbulent flow and $(c_3)$-fully developed rough turbulent flow

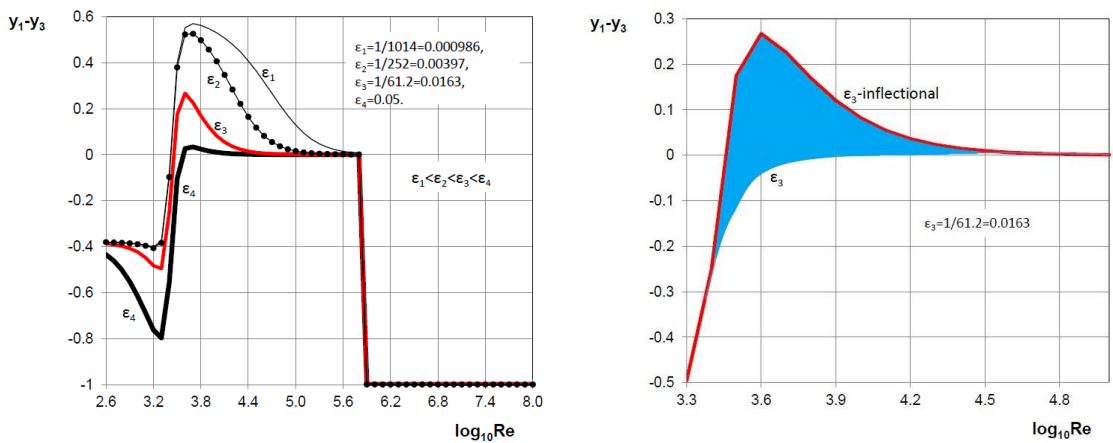

**Figure 7.** Function $y_1$-$y_3$ with the main purpose to provide Nikuradse's inflectional shape of the curves; on the right side: magnified detail from the left side – blue area presents the difference between Colebrook-White monotonic and Nikuradse's inflectional shape

*3.2. Friction factors*

Friction factors, which serve as inputs in the here presented unified formula, Eq. (1), will be here shown in more details; Here we present friction factors for $(a)$-laminar, $(c_1)$-smooth turbulent flow, and $(c_3)$-fully developed rough turbulent flow hydraulic regimes. Turbulent flow is a phenomena that still causes a stir among experts [29,30]. Thus, the here shown relation is suitable for flow of Newtonian fluids, whereas some restrictions are applied for gases [31].

*3.2.1. $(a)$-laminar*

Due to the overriding effect of viscosity forces in laminar flow, even rough inner pipe surface appears to be hydraulically smooth for the Reynolds numbers lower than about 3000,. Consequently, the roughness of walls, unless it is very significant, does not affect the flow resistance. Under these conditions of the flow, the friction coefficient $\lambda$ is always a function of the Reynolds number Re alone; Eq (10):

$$(a) \sim \lambda = \frac{64}{Re} \tag{10}$$

*3.2.2. $(c_1)$-smooth turbulent flow*

In the hydraulically smooth regime the friction factor, $\lambda$ is a function of the Reynolds number Re only and the resistance to flow is independent of the relative roughness of inner pipe surface ε. This regime is restricted to the relatively small values of the Reynolds number Re where the roughness of inner pipe surface is completely hidden in the laminar boundary sub-layer. The literature on the hydraulically smooth regime abounds with reliable friction factor equations. In generally, with some extensions and modifications, there are Blasius form or power law relationships; Eq. (11)-up, and von Karman-Prandtl form or logarithmic relationships; Eq. (11)-down:

$$\left. \begin{array}{l} (c_1) \sim \lambda = A \cdot Re^{-B} \\ (c_1) \sim \frac{1}{\sqrt{\lambda}} = C \cdot log_{10}(Re \cdot \sqrt{\lambda}) - D \end{array} \right\} \tag{11}$$

Some possible values for coeficient A and exponent B for the Blasius form or power law relationships [17,26,31]; Eq. (11)-up, is given in Table 1:

Table 1. Power law relations for the hydraulically smooth turbulent regime

| Equation in form: $\lambda = A \cdot Re^{-B}$ | Coefficient A | Exponent B |
|---|---|---|
| Renouard | 0.172 | 0.18 |
| 1/10 power law | 0.139 | 0.18 |
| modified 1/9 power law | 0.184 | 0.2 |
| 1/9 power law | 0.1748 | 0.2 |
| 1/8 power law | 0.2252 | 0.22 |
| 1/7 power law | 0.3052 | 0.25 |
| Müller | 0.3564 | 0.26 |
| Blasius | 0.3164 | 0.25 |
| Panhandle A | 0.08475 | 0.1461 |
| Panhandle B | 0.01471 | 0.03922 |
| IGT (Institute of Gas Technology) | 0.18086 | 0.19726 |
| Towler and Pope | 0.09458 | 0.15174 |
| Mokhatab | 0.02 | 0.185 |
| Hodanovič | 0.22 | 0.185 |

Regarding von Karman-Prandtl form or logarithmic relationships here is shown its basic form as used for a smooth part of the Colebrook equation, Eq. (12)-up. This formula is an updated version given by McKeon et al. [32], Eq. (12)-down:

$$\left.\begin{array}{l}(c_1) \sim \frac{1}{\sqrt{\lambda}} = 2 \cdot log_{10}(Re \cdot \sqrt{\lambda}) - 0.8 = -2 \cdot log_{10}\left(\frac{2.51}{Re} \cdot \frac{1}{\sqrt{\lambda}}\right) \\ (c_1) \sim \frac{1}{\sqrt{\lambda}} = 1.884 \cdot log_{10}(Re \cdot \sqrt{\lambda}) - 0.331 \end{array}\right\} \quad (12)$$

In addition, for the hydraulically smooth turbulent regime, among other, it is possible to use, Eq. (13), up to down: Genereaux, Leese, Nikuradse, Hermann, White and Konakov [17,26,31].

$$\left.\begin{array}{l}(c_1) \sim \frac{1}{\sqrt{\lambda}} = 1.6 \cdot log_{10}\left(\frac{Re \cdot \sqrt{\lambda}}{2}\right) - 0.6 \\ (c_1) \sim \lambda = 0.0072 + \frac{0.612}{Re^{0.35}} \\ (c_1) \sim \lambda = 0.0032 + \frac{0.221}{Re^{0.237}} \\ (c_1) \sim \lambda = 0.0054 + \frac{0.936}{Re^{0.3}} \\ (c_1) \sim \lambda = \frac{1.02}{(log_{10}(Re))^{2.5}} \\ (c_1) \sim \lambda = \frac{1}{(1.81 \cdot log_{10}(Re) - 1.5)^2} \end{array}\right\} \quad (13)$$

Though the Blasius forms or the power-law relationships, Eq. (11)-up, have the merit of simplicity, they also have certain disadvantages, one of which being that the relations can only be applied over a limited range of hydraulically smooth regime. Extrapolations beyond this range cannot be made with confidence. Figure 8 gives a comparison of some of the friction coefficients $\lambda$ used in the hydraulically smooth turbulent regime. As can be seen, the friction factor $\lambda$ corresponding to the Hodanovič equation has higher values than those for Panhandle B, see Table 1. This fact is a demonstration of how limited is the range of application of several of the available equations as many of them were developed for particular situations. For example, Panhandle B is valid for large-diametere pipelines, while Renoard is suitable for distribution PVC pipeline networks in urban areas. In addition, main constraint for using von Karman-Prandtl form or logarithmic relationships; Eq. (11)-down is its implicit form in respect of the flow friction $\lambda$.

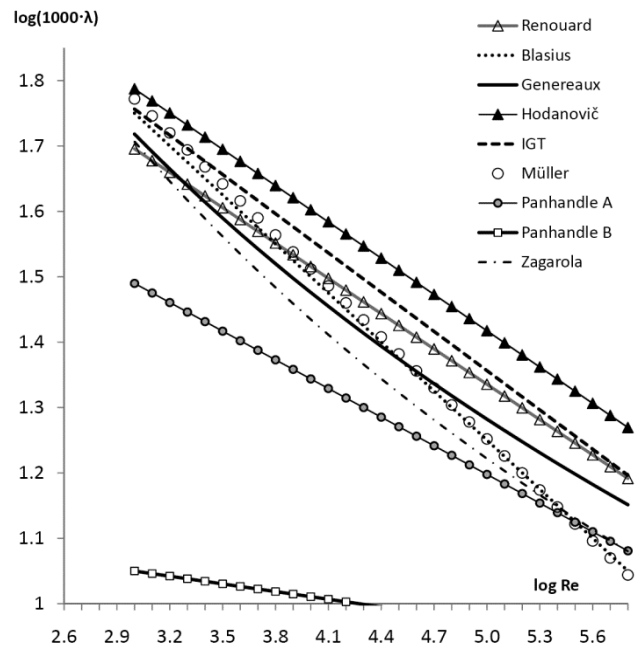

**Figure 8.** A comparison of some correlations for hydraulically smooth turbulent regime

*3.2.3. $(c_3)$-fully developed rough turbulent flow*

At high values of the Reynolds number Re, the friction factor $\lambda$ becomes a constant for a given relative roughness $\varepsilon$ [33]. In the case of the fully rough turbulent flow, the laminar sub-layer near the pipe wall practically does not exist and thus the flow is dominated by the relative roughness $\varepsilon$. The equation can be logarithmic; Eq. (14)-up, or power-law as used mostly in Russian engineering practice such as in so called Altshul or Shifrinson type of formulas; Eq. (14)-down [17]:

$$\left. \begin{aligned} (c_3) &\sim \frac{1}{\sqrt{\lambda}} = 1.14 - 2 \cdot log_{10}(\varepsilon) = -2 \cdot log_{10}\left(\frac{\varepsilon}{3.71}\right) \\ (c_3) &\sim \lambda = 0.11 \cdot \varepsilon^{0.25} \end{aligned} \right\} \quad (14)$$

A difference in results regarding Eq. (14) can be seen in Figure 2 of this paper.

*3.3. New unified flow friction formulation*

A novel formula for all hydraulic regimes is developed in order to easily encapsute separate formulas for different hydraulic regimes into the one coherent hydraulic regime. The structure of the proposed formula, $\lambda = (1 - y_1) \cdot (a) + (y_1 - y_3) \cdot (c_1) + y_2 \cdot (c_3)$, is based on using three switcing functions $y_1$, $y_2$ and $y_3$, and formulas for $(a)$-laminar, $(c_1)$-smooth turbulent flow, and $(c_3)$-fully developed rough turbulent flow. Consequently, all hydraulic regimes can be simulated in one simple formula including a sharp transition from laminar to the smooth turbulent $(b)$ and non-fully developed partially turbulent regime $(c_2)$. As already explained, the switching functions are set in such a way to produce always the Nikuradse's inflectional shape of the curves [34]. For example, two possible encapsulations in one unified coherent hydraulic model is given with Eq. (15) with the related diagrams in Figures 9 and 10, respectively.

$$\left. \begin{aligned} \lambda &= \overbrace{\frac{64}{Re}}^{(a)} \cdot (1 - y_1) + \overbrace{\frac{0.316}{Re^{0.25}}}^{(c_1)} \cdot (y_1 - y_3) + \overbrace{\frac{0.25}{log_{10}^2\left(\frac{\varepsilon}{3.71}\right)}}^{(c_3)} \cdot y_2 \\ \lambda &= \underbrace{\frac{64}{Re}}_{(a)} \cdot (1 - y_1) + \underbrace{\left(0.0032 + \frac{0.221}{Re^{0.237}}\right)}_{(c_1)} \cdot (y_1 - y_3) + \underbrace{0.11 \cdot \varepsilon^{0.25}}_{(c_3)} \cdot y_2 \end{aligned} \right\} \begin{matrix} I \\ II \end{matrix} \quad (15)$$

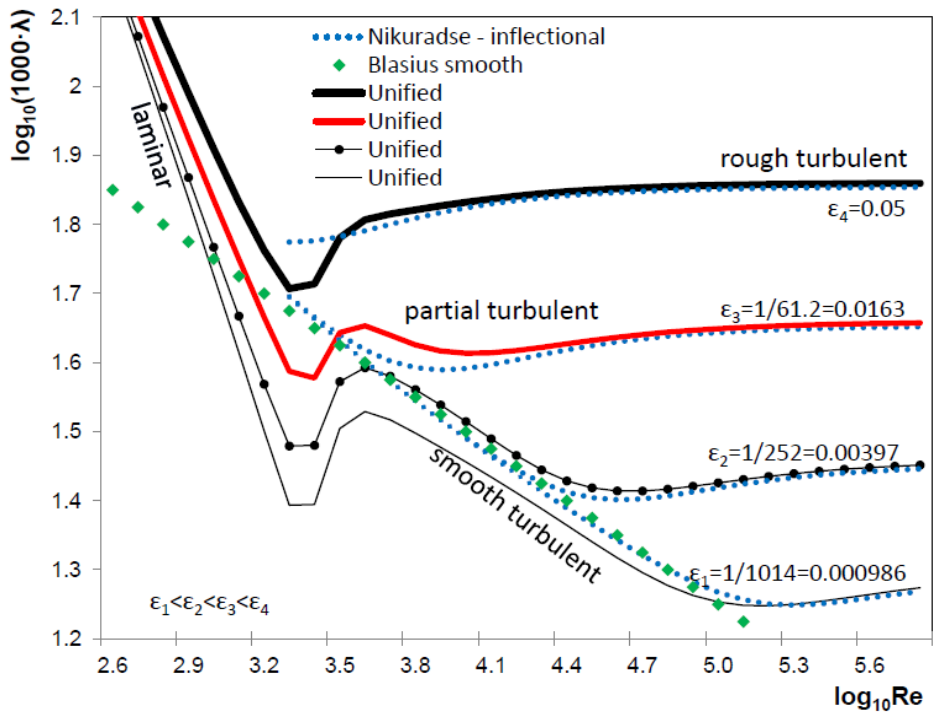

**Figure 9.** Unified hydraulic model *I*; Eq. (15)

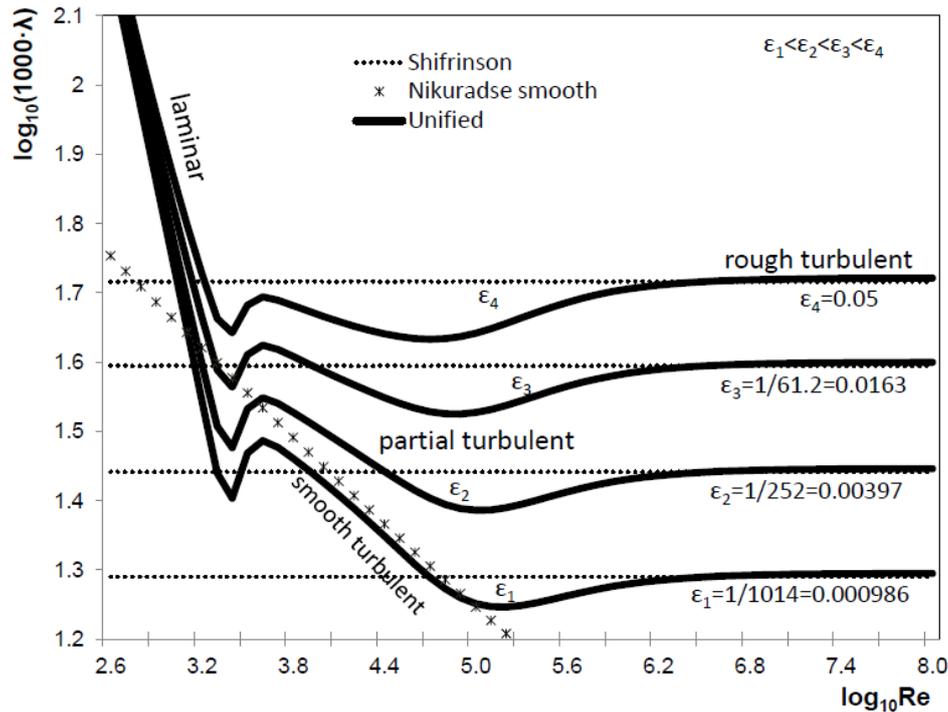

**Figure 10.** Unified hydraulic model *II*; Eq. (15)

In Figure 9, the blue dot-line shape is produces using the Colebrook equation with the extension for the Nikuradse's inflectional law; Eq. (5). Moreover, laminar flow $(a)$ is given as usually by $\lambda = \frac{64}{Re}$, whereas the hydraulically smooth turbulent regime $(c_1)$ with Blasius expression is represented as $\lambda = \frac{0.316}{Re^{0.25}}$. Finally, the fully developed turbulent regime $(c_3)$ with von Karman-Prandtl equation $\lambda = \frac{0.25}{log_{10}^2\left(\frac{\varepsilon}{3.71}\right)}$, is with the reference to Eq. (15)-*I*. For Figure 10, a similar approach is given by different formulas, as clearly indicated in Eq. (15)-*II*.

Using the same separate formulas for different hydraulic regimes from Eq. (15), one more additional encapsulation is possible, Eq. (16):

$$\left.\begin{array}{l} \lambda = \overbrace{\frac{64}{Re}}^{(a)} \cdot (1 - y_1) + \overbrace{\frac{0.316}{Re^{0.25}}}^{(c_1)} \cdot (y_1 - y_3) + \overbrace{0.11 \cdot \varepsilon^{0.25}}^{(c_3)} \cdot y_2 \\ \lambda = \underbrace{\frac{64}{Re}}_{(a)} \cdot (1 - y_1) + \underbrace{\left(0.0032 + \frac{0.221}{Re^{0.237}}\right)}_{(c_1)} \cdot (y_1 - y_3) + \underbrace{\frac{0.25}{log_{10}^2\left(\frac{\varepsilon}{3.71}\right)}}_{(c_3)} \cdot y_2 \end{array}\right\} \begin{array}{l} III \\ IV \end{array} \quad (16)$$

In such a way, after performing numerical tests, the most appropriate equation using separate available equations for different hydraulic can be encapsulated in a best way depending on particular circumstances required by each engineering project separately.

## 4. Conclusions

The paper presents new formula for a Newtonian fluids valid for all pipe flow regimes starting from laminar to the rough turbulent. Mentioned formula allows for the inflectional form what was presented in Nikuradse's experiment comparing to monotonic shape explained by Colebrook. The composition consists of switching functions and interchangeable formulas for laminar, smooth turbulent and rough turbulent flow.

A sudden failure of valves or other components, either related to hydraulic systems in civil and mechanical engineering [35-37], can cause also change of flow regime. Because of that it is very important to take into consideration also such cases. The here presented unified flow friction approach is flexible, as proposed equations for certain hydraulic flow regime can be easily changed. Although our previous experiences with artificial intelligence [38-40] showed that encapsulation of all flow friction regimes into a one coherent model is not a straightforward task, the here proposed form is simple. Thus, the here presented unified approach can be easily implemented in software codes. Moreover, as the proposed switching functions are carefully chosen in such a way to follow the Nikurdse's inflectional law of roughness, the here proposed unified approach seems to be more realistic compared with the classical implicitly given 80 years old Colebrook-White monotonic curves model [41-43]. The here presented switching functions are expressed by a simple rational functions and thus do not contain computationally expensive transcendental functions. Consequently, the here presented unified flow friction formulation has also a reasonable computational complexity.

**Author Contributions:** The paper is a product of the joint efforts of the authors who worked together on models of natural gas distribution networks. P.P. has scientific background in applied mathematics and programming while D.B.'s background is in control and applied computing in mechanical and petroleum engineering. Based on the idea of D.B., P.P. used numerical data provided by D.B. to develop the presented switching functions in HeuristicLab [computer software]. D.B. prepared figures and wrote the draft of this article.

**Conflicts of Interest:** The authors declare no conflict of interest. Neither the European Commission, VŠB—Technical University of Ostrava nor any person acting on behalf of them is responsible for the use which might be made of this publication.

**Nomenclature**

The following symbols are used in this paper:

| | |
|---|---|
| $\lambda$ | Darcy friction factor (Moody, Darcy–Weisbach or Colebrook); dimensionless |
| Re | Reynolds number; dimensionless |
| $\varepsilon$ | relative roughness of inner pipe surface; dimensionless |
| $(a)$ | laminar |
| $(c_1)$ | smooth turbulent non |
| $(c_2)$ | non-fully developed partially turbulent |
| $(c_3)$ | fully developed rough turbulent |
| $y_1, y_2$ and $y_3$. | switching functions |
| $e$ | exponential function |
| log | logarithmic function |
| ln | Napier natural logarithm |
| A, B, C, D and E | auxiliary terms |

```
```